\documentclass[12pt]{article}

\usepackage{amsmath,amssymb}
\usepackage{graphicx}% Include figure files
\usepackage{caption}
\usepackage{color}% Include colors for document elements
\usepackage{dcolumn}% Align table columns on decimal point
\usepackage{bm}% bold math
\usepackage[authoryear, round,sort]{natbib}
\usepackage{float}
\usepackage{hyperref}
\hypersetup{colorlinks = true, citecolor = blue, urlcolor = blue}
\newcommand{\R}{\mbox{$\mathbb{R}$}}

\let\hat\widehat
\let\tilde\widetilde

\definecolor{background-color}{gray}{0.98}

\title{Modal Regression using Kernel Density Estimation: a Review}
\author{Yen-Chi Chen\thanks{Department of Statistics, University of Washington}}
\date{}
\begin{document}
\maketitle

\begin{center}
%\subsubsection*{\small Article Type:}
%Advanced Review
%The Article Type denotes the intended level of readership for your article. An Editor may have mentioned a specific Article Type in your invitation letter; if so, please let them know if you think a different Article Type better suits your topic.

\hfill \break
\thanks

%\subsubsection*{Abstract}
%\begin{flushleft}
\begin{abstract}
We review recent advances in modal regression studies using kernel density estimation.
Modal regression is an alternative approach for investigating relationship between
a response variable and its covariates. 
Specifically, modal regression 
summarizes the interactions between the response variable and covariates
using 
the conditional mode or local modes.
%In the case of conditional mode, we call it uni-modal regression
%and in the case of conditional local modes, we call it multi-modal regression.
We first describe the underlying model of modal regression and its estimators based on kernel density estimation. 
We then review the asymptotic properties of the estimators and strategies for
choosing the smoothing bandwidth. 
We also discuss useful algorithms and similar alternative approaches for modal regression, and propose future direction in this field.
\end{abstract}
%\end{flushleft}
\end{center}

%\clearpage

% makes references listed with 1., 2., etc.  
%\makeatletter
%\renewcommand\@biblabel[1]{#1.}
%\makeatother
%\bibliographystyle{apsrev}

\renewcommand{\baselinestretch}{1.5}
\normalsize

%\clearpage

%\section*{\sffamily \Large GRAPHICAL TABLE OF CONTENTS} 
\section{Introduction} 

%\section*{\sffamily \Large INTRODUCTION} 
%\section*{\sffamily \Large First-order heading}
%\subsection*{\sffamily \large Second-order heading}
%\subsubsection*{\sffamily \normalsize Third-order heading}

%\section{Uni-Modal Regression}
%
%\subsection{Parametric approach}
%
%\subsection{Nonparametric approach}
%
%\subsection{Bayesian approach}

Modal regression is an approach for studying the relationship between a response variable $Y$ and its covariates $X$.
Instead of seeking the conditional mean, the modal regression
searches for the conditional modes \citep{sager1982maximum,collomb1986note,Lee1989} or 
local modes \citep{Einbeck2006,chen2016nonparametric} of the response variable $Y$ given the covariate $X=x$.
%Modal regression has been applied to various problems including analyzing AlzheimerÕs Disease \citep{wang2017cognitive}, 
The modal regression would be a more reasonable modeling approach
than the usual regression in two scenarios. 
First, when the conditional density function is skewed or has a heavy tail. 
When the conditional density function has skewness, the conditional mean may not provide a good
representation for summarizing the relations between the response and the covariate ($X$-$Y$ relation). 
The other scenario is when the conditional density function
has multiple local modes. 
This occurs when the $X$-$Y$ relation contains multiple patterns.
The conditional mean may not capture any of these patterns so it can be a
very bad summary; see, e.g., \cite{chen2016nonparametric} for an example. 
This situation has already been pointed out 
in \cite{tarter1993model}, where the authors argue that we should
not stick to a single function for summarizing the $X$-$Y$ relation
and they recommend looking for the conditional local modes.

Modal regression has been applied
to various problems such as 
predicting Alzheimer's disease \citep{wang2017cognitive}, 
analyzing dietary data \citep{zhou2016nonparametric}, 
predicting temperature \citep{Hyndman1996},
analyzing electricity consumption \citep{chaouch2017rate}, 
and studying the pattern of forest fire \citep{yao2014new}. 
In particular, \cite{wang2017cognitive}
argued that the neuroimaging features and cognitive assessment
are often heavy-tailed and skewed.
A traditional regression approach
may not work well in this scenario, so
the authors propose to use a regularized modal regression for predicting Alzheimer's disease.

%Modal regression has been used in transportation
%\citep{Einbeck2006}, astronomy \citep{Rojas2005}, meteorology
%\citep{Hyndman1996} and economics
%\citep{Huang2012jasa,Huang2013jasa}.  

The concept of modal regression was proposed in \cite{sager1982maximum}.
In this pioneering work, the authors stipulated that the conditional (global) mode be 
a monotone function of the covariate. 
\cite{sager1982maximum} also pointed out that 
a modal regression estimator can be constructed using a plug-in from a density estimate. 
\cite{Lee1989} proposed a linear modal regression that combined a smoothed 0-1 loss with a maximum likelihood estimator
(see equation \eqref{eq::argmin} for how these two ideas are connected). 
The idea proposed in \cite{Lee1989} was subsequently modified in many studies; see, e.g.,
\cite{Lee1989,manski1991regression,lee1993quadratic,lee1998semiparametric,kemp2012regression,yao2014new,krief2017semi}.

The idea of using conditional local modes has been pointed out in \cite{tarter1993model}
and the 1992 version of Dr. David Scott's book \emph{Multivariate density estimation: theory, practice, and visualization} \citep{scott1992multivariate}. 
The first systematic analysis
was done in \cite{Einbeck2006},
where the authors proposed a plug-in estimator using a kernel density estimator (KDE)
and computed their estimator by a computational approach modified from the meanshift algorithm
\citep{comaniciu2002mean,fukunaga1975estimation,cheng1995mean}.
The theoretical analysis and several extensions, including confidence sets, prediction sets,
and regression clustering were later studied in \cite{chen2016nonparametric}. 
Recently, \cite{zhou2016nonparametric} extended this idea to measurement error problems. 

The remainder of this review paper is organized as follows. 
In Section~\ref{sec::MR}, we formally define the modal regression model and discuss
its estimator by KDE. 
In Section~\ref{sec::thm}, we review the asymptotic theory of the modal regression estimators. 
Possible strategies for selecting the smoothing bandwidth 
and computational techniques
are proposed in Section~\ref{sec::bw} and \ref{sec::comp}, respectively.
In Section~\ref{sec::similar}, we discuss two alternative but similar approaches to modal regression -- the mixture of regression
and the regression quantization method.
The review concludes with some possible future directions in Section~\ref{sec::future}.

\section{Modal Regression}	\label{sec::MR}

For simplicity, we assume that the covariate $X$ is univariate with a compactly supported denisty function.
%We will call the conditional modes/local modes as the \emph{modal function}. 
Two types of modal regression have been studied in the literature. 
The first type, focusing on the conditional (global) mode, 
is called uni-modal regression \citep{sager1982maximum, collomb1986note,Lee1989,manski1991regression}.
%because it focuses on the global mode, often a single point at a given point $x$. 
The other type, which finds the conditional local modes, is calld
multi-modal regression \citep{Einbeck2006,chen2016nonparametric}.

\begin{figure}
\includegraphics[width=3in]{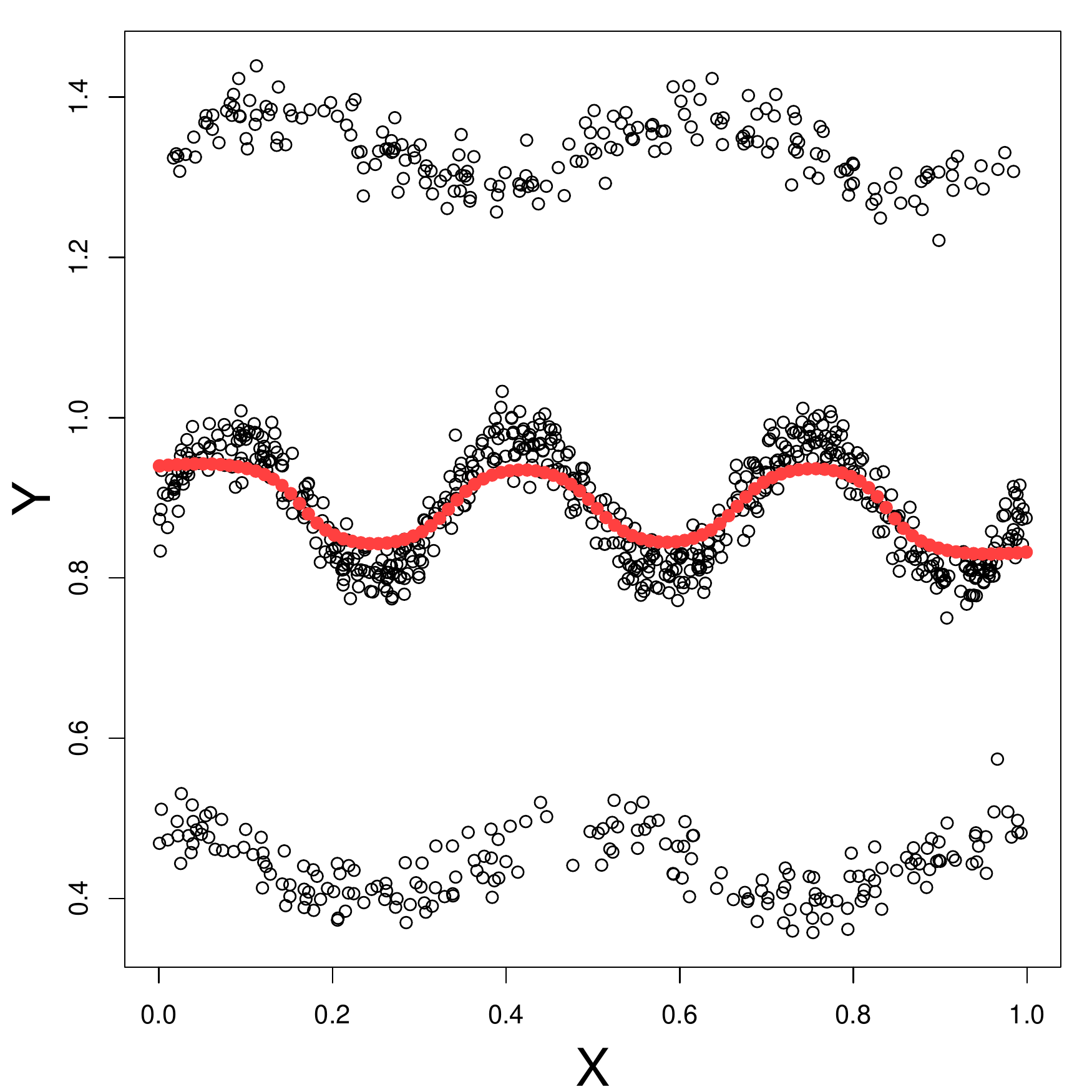}
\includegraphics[width=3in]{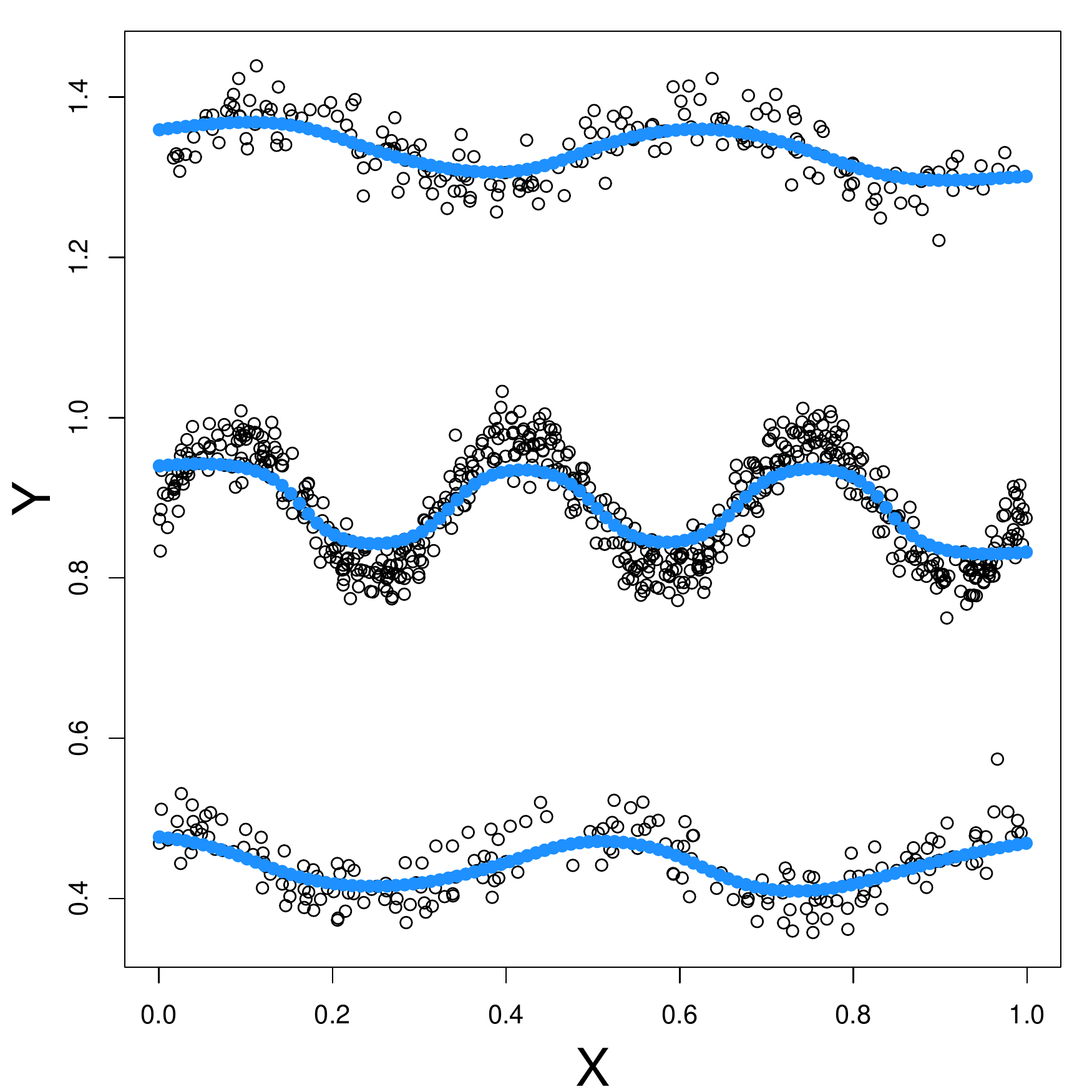}
\caption{Uni-modal regression (left; red curve) and multi-modal regression (right; blue curves)on a simulation
dataset with three components.
%Both regressions were estimated with a KDE.
%We estimate both the uni-modal and multi-modal regression using the KDE.
%In the left panel, we present the uni-modal regression (red curve). 
%In the right panel, we display the multi-modal regression (blue curve).
}
\label{fig::ex01}
\end{figure}

More formally, let $q(z)$ denote the probability density function (PDF) of a random variable $Z$. 
We define the operators
$$
{\sf UniMode}(Z) = \underset{z}{{\sf argmax}}\,\,\, q(z)
$$
and 
$$
{\sf MultiMode}(Z) = \{z: q'(z)=0, q''(z)<0\},
$$
which return the global mode and local modes of the PDF of $Z$, respectively.
Note that we need $q$ to be twice differentiable.
Uni-modal regression searches for the function
\begin{equation}
m(x)= {\sf UniMode}(Y|X=x) = \underset{y}{{\sf argmax}}\,\,\, p(y|x)
\label{eq::unimodal}
\end{equation}
whereas multi-modal regression targets 
\begin{equation}
M(x)= {\sf MultiMode}(Y|X=x) = \left\{y: \frac{\partial}{\partial y}p(y|x)=0,\frac{\partial^2}{\partial y^2} p(y|x)<0\right\}.
\label{eq::modal}
\end{equation}
Note that the modal function $M(x)$ may be a multi-valued function. Namely,
$M(x)$ may take multiple values at a given point $x$.
Figure~\ref{fig::ex01} presents examples of uni-modal and multi-modal regression
using a plug-in estimate from a KDE.

Because $p(y|x)= \frac{p(x,y)}{p(x)}$, 
the mode or local modes of $p(y|x)$ and $p(x,y)$ are equal for a given fixed $x$.
Thus, provided that $p(x)>0$,
we can rewrite both uni-modal and multi-modal regression in the following form: 
\begin{equation}
m(x)=\underset{y}{{\sf argmax}}\,\,\, p(x,y),\quad
M(x)= \left\{y: \frac{\partial}{\partial y}p(x,y)=0,\frac{\partial^2}{\partial y^2} p(x,y)<0\right\}.
\label{eq::modal2}
\end{equation}
That is, both types of modal regressions can be directly defined
through the joint PDF. 
Therefore, an estimated joint PDF can be inverted into a modal regression estimate.
Note that there are also Bayesian methods for modal regression; see, e.g., \cite{damien2017bayesian}.

% as long as we can estimate the joint PDF, we can invert the joint PDF estimate into
%an estimate of the modal regression.

\subsection{Estimating Uni-modal Regression}

%A simple approach of estimating the uni-modal regression is
%via the KDE \citep{sager1982maximum,collomb1986note,yao2012local}.
The KDE provides a simple approach for estimating uni-modal regression \citep{sager1982maximum,collomb1986note,yao2012local}.
After estimating the joint PDF,
we form a plug-in estimate for the uni-modal regression using KDE.
%then use the KDE to form a plug-in estimate for the uni-modal regression.
In more detail, let
\begin{equation}
\hat{p}_n(x,y) = \frac{1}{nh_1h_2}\sum_{i=1}^n K_1\left(\frac{X_i-x}{h_1}\right)K_2\left(\frac{Y_i-y}{h_2}\right)
\label{eq::KDE}
\end{equation}
be the KDE
where $K_1$ and $K_2$ are kernel functions such as Gaussian functions
and $h_1,h_2>0$ are smoothing parameters that control the amount of smoothing.
%Note that here we use a product kernel [xxx] (product of the two kernel functions) in estimating the joint PDF
%but one can use other approach such as a radial basis kernel. 
An estimator of $m$ is 
\begin{equation}
\hat{m}_n(x) =\underset{y}{{\sf argmax}}\,\,\, \hat{p}_n(x,y).
\label{eq::uni_modal}
\end{equation}
Note that the joint PDF can be estimated by other approaches such as
local polynomial estimation as well \citep{fan1996estimation,fan2004crossvalidation,Einbeck2006}.

Equation \eqref{eq::KDE} has been generalized to the case of
censored response variables.
\cite{ould2005strong,khardani2010some,khardani2011uniform}. 
%have generalized equation \eqref{eq::KDE}
%to the case where the response variable is censored. 
Suppose that instead of observing the response variables $Y_1,\cdots,Y_n$, we observe
$T_i = \min\{Y_i,C_i\}$ and an indicator $\delta_i=I(T_i=Y_i)$ that informs whether $Y_i$ is observed or not
and $C_i$ is an random variable that is independent of $X_i$ and $Y_i$. 
In this case, equation \eqref{eq::KDE} can be modified to
\begin{equation}
\hat{p}^\dagger_n(x,y) = \frac{1}{nh_1h_2}\sum_{i=1}^n K_1\left(\frac{X_i-x}{h_1}\right)K_2\left(\frac{T_i-y}{h_2}\right)\times \frac{\delta_i}{\hat{S}_n(T_i)},
\label{eq::censor}
\end{equation}
where $\hat{S}_n(t) $ is the Kaplan-Meier estimator \citep{kaplan1958nonparametric}
$$
\hat{S}_n(t)=\begin{cases}
\prod_{i=1}^n \left(1-\frac{\delta_{(i)}}{n-i+1}\right)^{I(T_{(i)}\leq t)}\quad &\mbox{if } t<T_{(n)},\\
0 &\mbox{otherwise},
\end{cases}
$$
with $T_{(1)}\leq T_{(2)}\leq \cdots \leq T_{(n)}$ being the ordered $T_i$'s
and $\delta_{(i)}$ being the value of $\delta$ for the $i$-th ordered observation.
Replacing $\hat{p}_n$ by $\hat{p}^\dagger_n$ in equation \eqref{eq::uni_modal}, 
we obtain a uni-modal regression estimator in the censoring case. 

%We can combine the KDE and a parametric model to estimate the uni-modal regression as well.
Uni-modal regression may be estimated parametrically as well.
%A key ingredient is that 
When $K_2$ is a spherical (box) kernel $K_2(x) =\frac{1}{2} I(|x|\leq 1)$, 
the {\sf argmax} operation is equivalent to the {\sf argmin} opeartor
on a flattened $0-1$ loss. 
In more detail, consider a 1D toy example with observations $Z_1,\cdots,Z_n$
and a corresponding KDE $\hat{q}(z)=\frac{1}{2nh}\sum_{i=1}^nI(|z-Z_i|\leq h)$ obtained with a spherical kernel. 
It is easily seen that
\begin{equation}
\begin{aligned}
\underset{z}{{\sf argmax}}\,\,\, \hat{q}(z)& =
\underset{z}{{\sf argmax}}\,\,\, \frac{1}{2nh}\sum_{i=1}^nI(|z-Z_i|\leq h) \\
&= \underset{z}{{\sf argmax}}\,\,\,\sum_{i=1}^nI(|z-Z_i|\leq h)\\
& = \underset{z}{{\sf argmin}}\,\,\, \sum_{i=1}^nI(|z-Z_i|> h).
\end{aligned}
\label{eq::argmin}
\end{equation}
Parametric uni-modal regression forms estimators using equation \eqref{eq::argmin}
or its generalizations
\citep{Lee1989,manski1991regression,lee1993quadratic,lee1998semiparametric,kemp2012regression,yao2014new,krief2017semi,khardani2017non}.
Parameters estimated through the maximizing criterion in equation \eqref{eq::argmin}
is equivalent to maximum likelihood estimation. 
Conversely, parameter estimation through the minimization procedure 
in equation \eqref{eq::argmin}
is equivalent to 
empirical risk minimization.
For example, to fit a linear model to $m(x) = \beta_0+\beta_1x$ \citep{Lee1989,yao2014new}, we can
use the fitted parameters
\begin{equation}
\begin{aligned}
\hat{\beta}_0, \hat{\beta}_1 
& = \underset{\beta_0,\beta_1}{{\sf argmax}}\,\, \frac{1}{2nh}\sum_{i=1}^n I(|\beta_0+\beta_1X_i-Y_i|\leq h)\\
&= \underset{\beta_0,\beta_1}{{\sf argmin}}\,\, \sum_{i=1}^n I(|\beta_0+\beta_1X_i-Y_i|> h)
\end{aligned}
\label{eq::linear_modal}
\end{equation}
to construct our final estimate of $m(x)$.

Using equation \eqref{eq::argmin}, we can always convert the problem of
finding the uni-modal regression into a problem of minimizing a loss function. 
Here, the tuning parameter $h$ can be interpreted as the smoothing bandwidth
of the applied spherical kernel.
Choosing $h $ is a persistently difficult task.
Some possible approaches will be discussed in Section~\ref{sec::bw}.

%
%This is a popular way to model the structure since we can 
%write the estimation procedure/population quantity as an optimization problem. 
%
%minimizing an interval procedure \citep{Lee1989} is equivalent to
%maximizing a `density' procedure using a spherical kernel with that width of interval \citep{yao2012local}
%
%
%Note that the limitation of uni-modal regression (estimating the conditional \emph{global} mode)
%cannot be used to capture the multiple structures within the data.

\subsection{Estimating Multi-modal Regression}
%A simple and classical approach of estimating $m$ or $M$ it to estimate the joint PDF
%via a kernel density estimation (KDE).
%We first estimate the joint PDF and then insert the estimated PDF into equation \eqref{eq::modal}
%to obtain our final estimator. 
%In more detail, let
%$$
%\hat{p}_n(x,y) = \frac{1}{nh_1h_2}\sum_{i=1}^n K_1\left(\frac{X_i-x}{h_1}\right)K_2\left(\frac{Y_i-y}{h_2}\right)
%$$
%be the KDE
%where $K_1,K_2$ are kernel functions such as Gaussian
%and $h_1,h_2>0$ are smoothing parameters.
%Note that here we use a product kernel [xxx] (product of the two kernel functions) in estimating the joint PDF
%but one can use other approach such as a radial basis kernel. 

%In the case of a multi-modal regression,
Like uni-modal regression,
multi-modal regression can be estimated using a plug-in estimate from the KDE \citep{Einbeck2006,chen2016nonparametric}.
Recalling that $\hat{p}_n(x,y)$ is the KDE of the joint PDF,
an estimator of $M(x)$ is
\begin{equation}
\hat{M}_n(x)= \left\{y: \frac{\partial}{\partial y}\hat{p}_n(x,y)=0,\frac{\partial^2}{\partial y^2} \hat{p}_n(x,y)<0\right\}.
\label{eq::modal_est}
\end{equation}
%Essentially, equation \eqref{eq::modal_est} is a plug-in estimate of equation \eqref{eq::modal} using
%the KDE. 
Namely, we use the conditional local modes of the KDE
to estimate the conditional local modes of the joint PDF. 
Plug-ins from a KDE have been applied in estimations of
many structures \citep{scott2015multivariate, Chen_2017} such as the regression function\citep{nadaraya1964estimating,watson1964smooth}, 
modes \citep{Chacon2012,chen2016comprehensive}, ridges \citep{genovese2014nonparametric,chen2016nonparametric}, and level sets \citep{rinaldo2010generalized,chen2017density}.
An alternative way of estimating the multi-modal regression was proposed in \cite{sasaki2016modal}.

In the measurement error case where the covariates $X_1,\cdots,X_n$ are observed with noises,
we can replace $K_1$ by a deconvolution kernel to obtain a consistent estimator \citep{zhou2016nonparametric}. 
In more detail, let 
$$
W_i = X_i+U_i,\,\, i=1,\cdots, n,
$$
where $U_1,\cdots,U_n$ are IID measurement errors that are independent of the covariates and responses.
We assume that the PDF of $U_1$, $f_U(u)$, is known.
Here we observe not
$X_i$'s but pairs of $(W_1,Y_1),\cdots,(W_n,Y_n)$. 
Namely, we observe the response variable and its corrupted covariate.
In this case, \eqref{eq::KDE} is replaced by
\begin{equation}
\tilde{p}_n(x,y) = \frac{1}{nh_1h_2}\sum_{i=1}^n K_{U}\left(\frac{W_i-x}{h_1}\right)K_2\left(\frac{Y_i-y}{h_2}\right),
\label{eq::de_KDE}
\end{equation}
where 
$$
K_U(t) = \frac{1}{2\pi} \int e^{-its}\frac{\phi_{K_1}(s)}{\phi_U(s/h_1)}ds
$$
with $\phi_{K_1}$ and $\phi_U$ being the Fourier transforms of $K_1$ and $f_U$, respectively.
The estimator of $M$ then becomes the conditional local modes of $\tilde{p}$:
\begin{equation}
\tilde{M}_n(x)= \left\{y: \frac{\partial}{\partial y}\tilde{p}_n(x,y)=0,\frac{\partial^2}{\partial y^2} \tilde{p}_n(x,y)<0\right\}
\label{eq::de_modal_est}
\end{equation}
For more details, the reader is referred to \cite{zhou2016nonparametric}.

\begin{figure}
\includegraphics[width=3in]{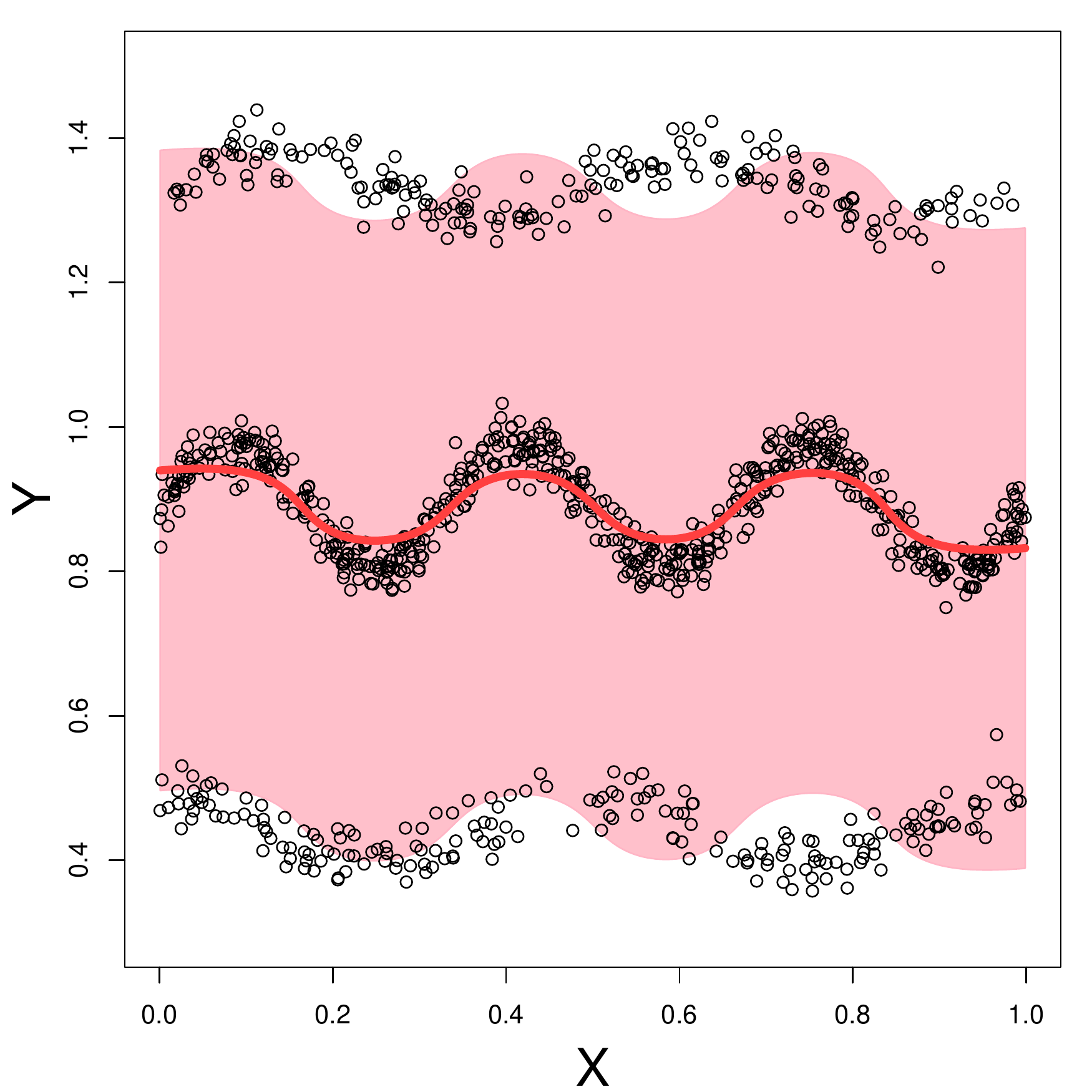}
\includegraphics[width=3in]{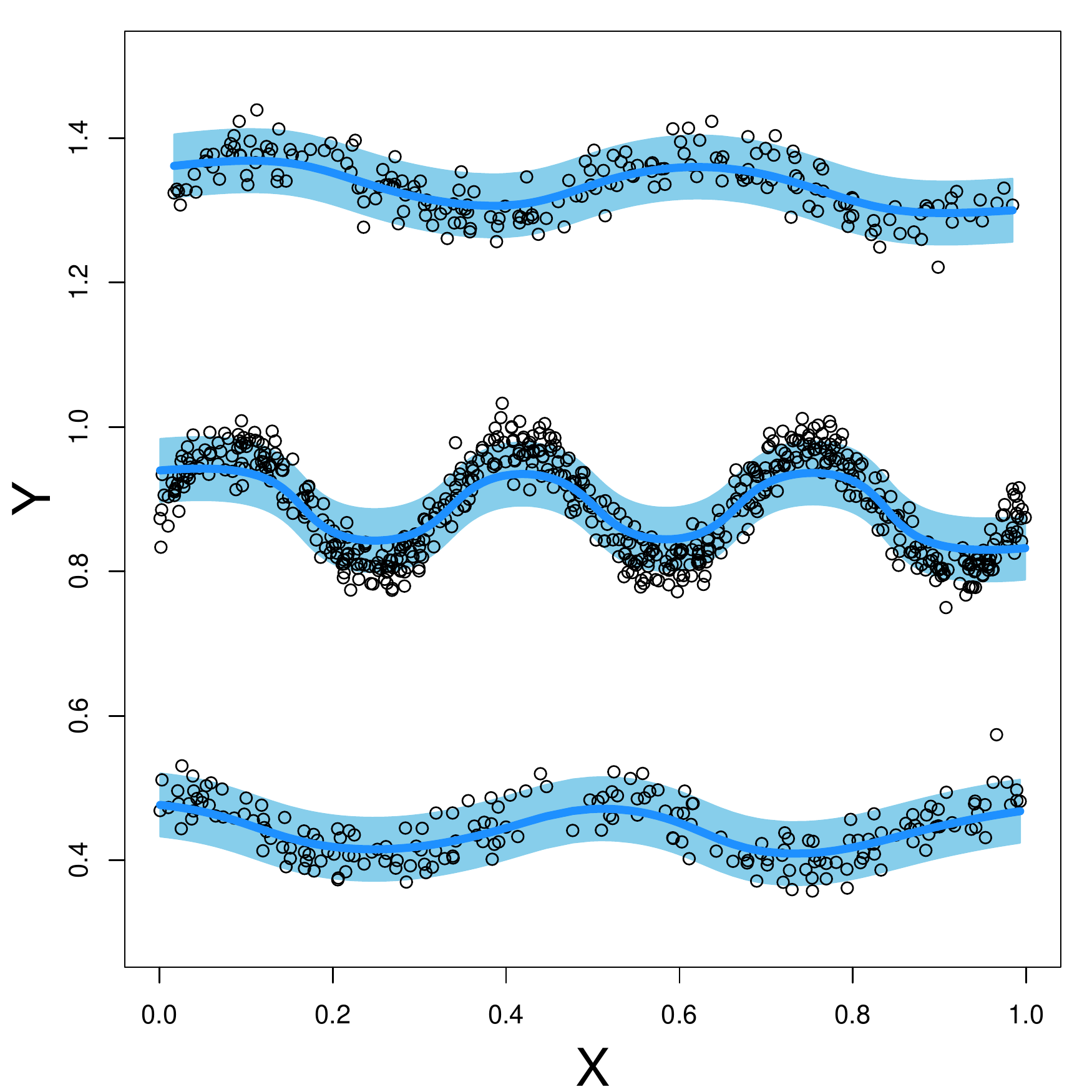}
\caption{90\% prediction regions constructed from uni-modal regression (pink area in the left panel) 
and multi-modal regression (light blue area in the right panel). 
%We construct a 90\% prediction region using both uni-modal regression and multi-modal regression.
Clearly, the prediction region is much smaller in
the multi-modal regression than in uni-modal regression because
multi-modal regression detects all components whereas
uni-modal regression discovers only the main component. 
}
\label{fig::ex02}
\end{figure}

\subsection{Uni-modal versus Multi-modal Regression}

Uni-modal and multi-modal regression have their own advantages and disadvantages. 
Uni-modal regression is an alternative approach for
summarizing the covariate-response relationship using
a single function. 
Multi-modal regression performs a similar job
but allows a multi-valued summary function.
When the relation between the response and the covariate is complicated or
has several distinct components (see, e.g., Figure \ref{fig::ex01}), 
multi-modal regression may detect the hidden relation that cannot be found by uni-modal regression.
In particular, the prediction regions tend to be smaller in multi-modal regression than in uni-modal regression
(see Figure~\ref{fig::ex02} for an example).
However, multi-modal regression often returns a multi-valued function that is 
more difficult to interpret than
the output from a uni-modal regression.

\section{Consistency of Modal Regression}	\label{sec::thm}

\subsection{Uni-Modal Regression}

Uni-modal regression often makes some smoothness assumptions\footnote{In \cite{Lee1989},
the assumptions are either symmetric and homogeneous errors or non-symmetric and heterogeneous error. } on the conditional 
density function $p(y|x)$ over variable $y$ \citep{Lee1989}.
These assumptions are made
to convert the mode hunting problem into a minimization or maximization problem
in equation \eqref{eq::argmin}.
%We can see the need of this assumption if we are going to use equation \eqref{eq::argmin}
%to convert the mode hunting problem into a loss/risk minimization problem. 
Equation \eqref{eq::argmin} implies that many estimators implicitly smooth the data by a spherical kernel 
then select the point that maximizes the result.
Thus, the estimator converges to the mode of a \emph{smoothed} density function (the expectation of the KDE). 
To ensure that the mode of the smoothed density function remains at the same location
as the mode of the original density function, a symmetric assumption is necessary.
%If we use a parametric model and a box kernel for variable $Y$, 
The convergence rate of a parametric model with a box kernel for variable $Y$
is $O_P(n^{-1/3})$ \citep{Lee1989}.
If the box kernel is replaced by a quadratic kernel, the convergence rate becomes $O_P(1/\sqrt{n})$ under suitable assumptions
\citep{lee1993quadratic}.
A nonparametric convergence rate 
was derived in
\cite{yao2014new} under a weaker assumption.

When estimating $m(x)$ using a
plug-in $\hat{m}_n(x)$ from KDE,
the convergence rate depends on the assumptions. 
If the conditional density possesses good characteristics (such as symmetry), then
\begin{equation}
\hat{m}_n(x)-m(x) = O(h_1^2) + O_P\left(\sqrt{\frac{1}{nh_1}}\right)
\label{eq::rate_NP0}
\end{equation}
when we are using the a Gaussian kernel or a first-order local polynomial estimator \citep{yao2012local}.
Besides the convergence rate, \cite{yao2012local} also derived the asymptotic normality of the estimator:
$$
\sqrt{nh_1}\left(\frac{\hat{m}_n(x)-m(x) - h_1^2 b(x)}{\sigma(x)}\right) \overset{D}{\rightarrow} N(0,1),
$$
where $b(x),\sigma(x)$ are functions describing the asymptotic bias and variance.
%and depending only on the population density $p(x,y)$ and the kernel functions. 
Note that the convergence rate and asymptotic normality are very similar to 
the usual nonparametric estimators. 
Under the assumptions on conditional density, the covariate is more responsible for
the smoothing effect than the response.
%The smoothing effect comes mostly from the covariate rather than from the 
%response
%due to the assumptions on the conditional density.

Various studies have reported the convergence of uni-modal regression with
dependent covariates
\citep{collomb1986note,ould1993estimation,ould1997note,ould2005strong,khardani2010some,
dabo2010note,khardani2011uniform,attaoui2014nonparametric}.
Strong consistency was investigated in \citep{collomb1986note,ould1993estimation,ould1997note}. 
The convergence rate has also been derived in uni-modal regression with functional dependent covariates
\cite{dabo2010note,attaoui2014nonparametric},
%considered the case of functional dependent covariates and derived the convergence rate.
and with censored response
\cite{ould2005strong,khardani2010some,khardani2011uniform}.

%functional dependent covariate \cite{dabo2010note,attaoui2014nonparametric}
%
%random censoring and dependent process \citep{khardani2010some,khardani2011uniform} usual nonparametric rate

%The main reason is that the location of mode is from applying an {\sf argmax} operation 
%to the conditional PDF.
%Since the estimated conditional PDF is a smooth function
%and has a nonparametric convergence rate that we are familiar with, 
%by the 
%If the conditional density is not 

%
%
%xxx\citep{yao2012local} has a faster rate but they require a bit stronger condition (nearly symmetric error)
%and they focus on estimating the conditional \emph{global} mode.
%
%Local polynomial estimate for conditional global mode

\subsection{Multi-Modal Regression}

Measuring the quality of modal regression is a difficult task
because the estimator $\hat{M}_n$ and the parameter of interest $M$
are both multi-valued functions.
$\hat{M}_n(x)$ and $M(x)$ are collections (sets) of values/points at each given point $x$.

We now define the Hausdoff distance, a popular measure of evaluating the difference 
between two sets. 
The Hausdorff distance between two sets For two given sets $A,B\subset \R^k$ is given by
\begin{align*}
{\sf Hausdorff}(A,B) &= \inf\{r\geq0: A\subset B\oplus r , B\subset A\oplus r\}\\ 
&= \max\left\{\sup_{x\in A} d(x,B), \sup_{x\in B}d(x,A)\right\},
\end{align*}
where $A\oplus r = \{x\in\R^k: d(x,A) \leq r\}$ is an augmented set of $A$ and 
$d(x,A) = \inf_{y\in A}\|x-y\|$ is the projection distance from point $x$ to set $A$.  
Specifically, the Hausdorff distance is the maximum projection distance between sets $A$ and $B$
and can be viewed as an $L_\infty$ distance of sets. 
As such, the Hausdorff distance has been applied as quality measure in estimating local modes \citep{chen2016comprehensive},
ridges \citep{genovese2014nonparametric}, and level sets \citep{chen2017density}, so it is excellently
suitable for measuring the distance between
$\hat{M}_n(x)$ and $M(x)$.

The \emph{pointwise error} at a given point $x$ is defined as
$$
\Delta_n(x) = {\sf Hausdorff}\left(\hat{M}_n(x), M(x)\right). 
$$
This pointwise error is similar to the usual pointwise error of estimating a regression function.
%that quantify the estimation error at point $x$.
Based on the pointwise error, we can easily define the \emph{mean integrated square error (MISE)}
and uniform errors
\begin{align*}
{\sf MISE}_n = \int \Delta^2_n(x)dx,\quad\quad
\Delta_n = \sup_x \Delta_n(x).
\end{align*}
These quantities are generalized from the errors
in nonparametric literature \citep{scott2015multivariate}.

The convergence rate of $\hat{M}_n(x)$ has been derived in \cite{chen2016nonparametric}:
\begin{equation}
\begin{aligned}
\Delta_n(x) &= O(h_1^2+h_2^2) + O_P\left(\sqrt{\frac{1}{nh_1h_2^3}}\right)\\
\Delta_n &= O(h_1^2+h_2^2) + O_P\left(\sqrt{\frac{\log n}{nh_1h_2^3}}\right)\\
{\sf MISE}_n &= O(h_1^4+h_2^4) + O\left(\frac{\log n}{nh_1h_2^3}\right).
\end{aligned}
\label{eq::rate_NP}
\end{equation}
The bias is now contributed by smoothing covariates and response. 
The convergence rate of the stochastic variation, $O_P\left(\sqrt{\frac{1}{nh_1h_2^3}}\right)$,
depends on the amount of smoothing in both the covariate and response variable as well.
The component $h_2^3$ can be decomposed as $h_2 \cdot h_2^2$, 
where the first part $h_2$ is the usual smoothing and the second part, $h_2^2$,
is from derivative estimations.
Note that the convergence rates in equation \eqref{eq::rate_NP} require no symmetric-like assumption
on the conditional density, but only smoothness and bounded curvature at each local mode. 
Therefore, the assumptions ensuring a consistent estimator are much weaker in
multi-modal regression than in uni-modal regression.
%we obtain a consistent estimator under a much weaker assumption. 
However, the convergence rate is much slower in multi-modal regression than in 
uni-modal regression (equation \eqref{eq::rate_NP0}).
%the price we pay in equation \eqref{eq::rate_NP} is the slow convergence rate -- 
%the convergence rate is slower than the one in equation \eqref{eq::rate_NP0}.
Note that under the same weak assumptions as multi-modal regression,
uni-modal regression can also be consistently estimated by a KDE
and the convergence rate will be the same as equation \eqref{eq::rate_NP}.

\cite{chen2016nonparametric} also derived the asymptotic distribution and a bootstrap theory of $\Delta_n$.
When we ignore the bias,
the uniform error converges to the maximum of a Gaussian process
and the distribution of this a maximum can be approximated 
by the empirical bootstrap \citep{Efron1979}. 
Therefore, by applying the bootstrap, one can construct a confidence band
for the modal regression.

In the case of measurement errors, a similar convergence rate to equation \eqref{eq::rate_NP}
can also be derived under suitable conditions \citep{zhou2016nonparametric}. 
Note that in this case, the distribution of measurement errors also affects the estimation quality.
%An ordinary smooth measurement error will lead to the same rate as equation \eqref{eq::rate_NP}.

\section{Bandwidth Selection}	\label{sec::bw}

Modal regression estimation often involves some tuning parameters.
In a parametric model, we have to choose a window size $h$ in equation \eqref{eq::argmin}. 
In other models, we often require two smoothing bandwidths: 
one for the response variable, the other for the covariate. 
Here we briefly summarize some bandwidth selectors
proposed in the literature.

\subsection{Plug-in Estimate}

In \cite{yao2012local}, uni-modal regression was estimated by a local polynomial estimator. 
One advantage of uni-modal regression is
the closed-form expression of the 
first-order error. 
Therefore, we can use a plug-in approach to obtain an initial error estimate and
convert it into a possible smoothing bandwidth.
This approach is very similar to the plug-in bandwidth selection in the 
density estimation problem \citep{sheather2004density}. 

This approach was designed to 
optimally estimating the error of a uni-modal regression estimate.  
%Later we will see that many other alternatives are designed for
%optimizing the quality of estimating the conditional density. 
However, this method is often not applicable
to multi-modal regression because 
a closed-form expression of the first-order error is often unavailable.  
Moreover, this approach requires a pilot estimate for the first error. 
If the pilot estimate is unreliable, the performance of the bandwidth selector may be seriously compromised.

\subsection{Adapting from Conditional Density Estimation}

%\cite{zhou2017bandwidth} 

A common approach for selecting the tuning parameter
is based on optimizing the estimation accuracy of conditional density function \citep{fan1996estimation,fan2004crossvalidation}. 
For instance,
the authors of \cite{Einbeck2006} 
adapted the smoothing bandwidth to
multi-modal regression 
by
optimizing the conditional density estimation rate and the Silverman's 
normal reference rule \citep{Silverman1986}.

The principle of adapting from estimating the conditional density often relies on
optimizing the integrated squared-errors:
%The density estimation CV is based on the integrated squared-error criterion:
\begin{align*}
ISE &= \int \int \left(\hat{p}_n(y|x)-p(y|x)\right)^2 p(x) \omega(x)dxdy\\
&=\int \int \hat{p}_n^2(y|x)p(x) \omega(x)dxdy\\ &- 2\int \int \hat{p}_n(y|x)p(y|x)p(x) \omega(x)dxdy + \int \int p^2(y|x)p(x) \omega(x)dxdy,
\end{align*}
where $\omega(x)$ is a user-selected weight function.
For simplicity,
one can choose $\omega(x) =1$ over the range of interest. 
Note that in density estimation literature, bandwidth selection by this expansion is called the CV criterion \citep{sheather2004density}. 

However, the ISE involves unknown quantities so it must be estimated. 
Depending on the estimating procedure, 
there are many other approaches such as
the
regression-based approach, bootstrap method,
and cross-validation approach; see
\cite{zhou2017bandwidth} for a comprehensive review.

Although this approach is simple and elegant, 
a good density estimator does not 
guarantee a good estimator of the local modes.
As is seen in equation \eqref{eq::rate_NP}, the convergence rate 
is actually slower when estimating local modes than when estimating density.

%Actually, the convergence rate is related to gradient estimation so an optimal rate
%in density estimation turns out to be an undersmoothing rate
%for estimating the local modes. 

\subsubsection{CV-SIMEX method}

The bandwidth selection method in
\cite{zhou2016nonparametric}, designed for measurement errors,
combines density estimation CV with simulation extrapolation (SIMEX; \citealt{cook1994simulation}).

Because the last quantity in the ISE is independent of the tuning parameter, and $p(x)dx = dF(x)$ and
$p(y|x) p(x)dxdy = dF(x,y)$ 
can be replaced by their empirical versions $d\hat{F}_n(x)$ and $d\hat{F}_n(x,y)$, 
the CV criterion can be estimated by
\begin{equation}
CV(h_1,h_2) =\frac{1}{n} \sum_{i=1}^n\int \hat{p}_{-i,n}^2(y|X_i)\omega(X_i)dy - \frac{2}{n}\sum_{i=1}^n \omega(X_i) \hat{p}_{-i,n}(Y_i|X_i),
\label{eq::CV}
\end{equation}
where $\hat{p}_{-i,n}(y|x)$ is the estimated conditional density without $i$-th observation (leave $i$-th observation out).
If the covariates $X_1,\cdots,X_n$ are known, we can choose $h_1$ and $h_2$ by minimizing equation \eqref{eq::CV}. 

Measurement errors manifest as noise in the corrupted covariates $W_1,\cdots,W_n$. 
In the CV-SIMEX approach \citep{zhou2016nonparametric}, $h_2$ is determined by Siverman's rule \citep{Silverman1986}
and we give a brief summary for the selection of $h_1$ as follows.
We first
generate $W^*_1,\cdots,W^*_n$ where $W^*_i = W_i + U^*_i$
and $U_1^*,\cdots,U^*_n$ are IID from the measurement error distribution.
Then we construct the estimator $\hat{p}^*_{n}$ by replacing $X_1,\cdots,X_n$ by $W^*_1,\cdots,W^*_n$. 
We modify equation \eqref{eq::CV} by
replacing $X_1,\cdots,X_n$ by $W_1,\cdots,W_n$
and replacing $\hat{p}_n$ by $\hat{p}^*_{n}$
Note that the weight $\omega$ will also be updated according to the range of $W$'s. 
Let $CV^*(h_1)$ be the resulting CV criterion.
%After computing $h_1^*$,
Now we compute another CV criterion as follows. 
We 
generate $W^{**}_1,\cdots,W^{**}_n$ where $W^{**}_i = W^*_i + U^{**}_i$
and $U_1^{**},\cdots,U^{**}_n$ are IID from the measurement error distribution.
Similar to the previous steps, we compute a new (conditional) density estimator $\hat{p}^{**}_n$ by replacing $X_1,\cdots,X_n$ by 
$W^{**}_1,\cdots,W^{**}_n$.
To obtain a new CV criterion, we again modify equation \eqref{eq::CV} by
replacing $X_1,\cdots,X_n$ by $W^{*}_1,\cdots,W^{*}_n$
and replacing $\hat{p}_n$ by $\hat{p}^{**}_{n}$. 
This leads to a new CV criterion which we denoted as $CV^{**}(h_1)$. 
We then repeat the above process multiple times
and calculate the average $\overline{CV}^*(h_1)$ and $\overline{CV}^{**}(h_1)$. 
Then we choose $h_1^* $ to be the minimizer of $\overline{CV}^*(h_1)$
and $h_1^{**}$ to be the minimizer of $\overline{CV}^{**}(h_1)$. 
The final choice of smoothing bandwidth is $\tilde{h}_1 = \frac{h_1^{*2}}{h_1^{**}}$.

%To obtain a CV value for each of $h_1$,
%we replace $X_1,\cdots,X_n$ in equation \eqref{eq::CV} by $W^*_1,\cdots,W^*_n$. 
%After iterating this procedure multiple times, we average the CV value for each $h_1$ 
%and select the smoothing bandwidth with the lowest CV value. 
%
%We repeat the above sampling process multiple times
%and take an average to obtain a modified CV criterion.
%By optimizing the equation modified CV criterion, 
%we obtain the first smoothing bandwidth $h_1^*$. 

%We denote this choice as $h_1^{**}$. 
%We first replace $X_1,\cdots,X_n$ in equation \eqref{eq::CV} by $W_1,\cdots,W_n$, and obtain the first smoothing bandwidth $h_1^*$
%by optimizing the equation. 
%We then
%generate $W^*_1,\cdots,W^*_n$ where $W^*_i = W_i + U^*_i$
%and $U_1^*,\cdots,U^*_n$ are IID from the measurement error distribution.
%To obtain a CV value for each of $h_1$,
%we replace $X_1,\cdots,X_n$ in equation \eqref{eq::CV} by $W^*_1,\cdots,W^*_n$. 
%After iterating this procedure multiple times, we average the CV value for each $h_1$ 
%and select the smoothing bandwidth with the lowest CV value. 
%We denote this choice as $h_1^{**}$. 
%The final choice of smoothing bandwidth is $\tilde{h}_1 = \frac{h_1^{*2}}{h_1^{**}}$. 

The CV procedure assesses the quality of estimating the conditional density. 
The simulation process (SIMEX part) exploits the similarity between the optimal $h_1$--$h_1^*$ relation
and the $h_1^*$--$h_1^{**}$ relation, i.e., $\frac{h_{1,opt}}{h_1^*} \approx \frac{h_1^*}{h_1^{**}}$. 
Thus, the smoothing bandwidth is selected by equating this approximation.

However,
CV-SIMEX optimizes
the quality of estimating the conditional density, not the
conditional local modes. 
As confirmed in equation \eqref{eq::rate_NP}, the optimal convergence rate
differs between density and local modes estimation, so this choice would undersmooth the local modes estimation.

\subsubsection{Modal CV-criteria}

\cite{zhou2017bandwidth} proposed a generalization of the density estimation CV criterion to
the multi-modal regression. 
The idea is to replace the ISE by 
$$
ISE_M = \int {\sf Hausdorff}^2\left(\hat{M}_n(x),M(x)\right) p(x) \omega(x)dx
$$
and find an estimator of the above quantity. Optimizing the corresponding
estimated $ISE_M$ leads to
a good rule for selecting the smoothing bandwidth.
In particular, \cite{zhou2017bandwidth} proposed 
to use a bootstrap approach to estimate $ISE_M$. 
The quantity $ISE_M$ is directly tailored to the modal regression rather than conditional density estimation
so it reflects the actual accuracy of modal regression.

\subsection{Prediction Band Approach}

Another bandwidth selector for multi-modal regression was proposed in
\cite{chen2016nonparametric}.
This approach optimizes the size of the prediction bands using a cross-validation (CV) in the regression analysis. 
%We first choose a prediction level such as $95\%$ and
After selecting a prediction level (e.g., 95\%), 
the data is split into a training set and a validation set. 
The modal regression estimator is constructed from the data in the training set,
and the residuals of the observations are derived from the validation set. 
The residual of a pair $X_{val}, Y_{val}$ is based on the shortest distance to the nearly conditional local mode.
Namely, the residual for an observation $(X_{val},Y_{val})$ in the validation set is $e_{val} = \min_{y\in \hat{M}_n(X_{val})} \|Y_{val} - y\|$. 
The 95\% quantile of the residuals specifies the radius of a 95\% prediction band. 
%The width of prediction bands will be bands with width 2 times the radius around each estimate conditional local modes
The width of a prediction band is twice the estimated radius.
After repeating this procedure several times as the usual CV,
we obtain an average size (volume of the prediction band) of the prediction band of each smoothing bandwidth.
%We then repeat this procedure as the usual cross-validation to obtain an average size of the prediction band
%of each smoothing bandwidth. 
The smoothing bandwidth is chosen to be the one that has the smallest (in terms of volume) prediction band. 

When $h$ is excessively small, 
there will be many conditional local modes which leads to
a large prediction band.
On the other hand, when $h$ is too large, the number of conditional local modes are small
but each of them has a very large band, yielding a large total size of the prediction band.
Consequently, this approach leads to a stable result. 

However, the prediction band approach is beset with several problems.
First, the optimal choice of smoothing bandwidth depends on the prediction level,
which cannot be definitely selected at present. 
Second, calculating the size of a band is computationally challenging 
in high dimensions. 
%And the overall computational cost is large since we need to perform the cross-validation. 
Third, there is no theoretical guarantee that the selected bandwidth follows
the optimal convergence rate.

Note that \cite{zhou2017bandwidth} proposed 
a modified CV criterion for approximating
the size of prediction band without specifying the prediction level. 
This approach avoids the problem of selecting a prediction level
and is computationally more feasible.

\section{Computational Methods}	\label{sec::comp}

In the modal regression estimation, a closed-form solution
to the estimator is often unavailable. 
Therefore, 
the estimator must be computed by a
numerical approach.
The parameters in uni-modal regression with a parametric model
can be estimated by a mode-hunting procedure \citep{Lee1989}. 
When estimating uni-modal regression by a nonparametric approach,
the conditional mode can be found by the gradient ascent method
or an EM-algorithm \citep{yao2012local,yao2014new}. 

The conditional local modes in
multi-modal regression can also be found by a gradient ascent approach.
If the kerne function of the response variable $K_2$ has a nice form such 
as being a Gaussian function, gradient ascent can be easily performed by
a simple algorithm called meanshift algorithm \citep{comaniciu2002mean,fukunaga1975estimation,cheng1995mean}.
In the following, we briefly review the EM algorithm and the meanshift algorithm
for finding modes
and explain their applications to modal regression.
%\subsection{Mode Estimation}

%\section{Computational Method}

\subsection{EM Algorithm}

A common approach for finding uni-modal regression
is the EM algorithm \citep{dempster1977maximum, wu1983convergence}.  
In the case of modal regression,
we use the idea from a modified method called the modal EM algorithm \citep{Li2007,yao2012local}.
For simplicity, we illustrate the EM algorithm using 
the linear uni-modal regression problem  
with a single covariate \citep{yao2014new}. 
Let $(X_1,Y_1),\cdots,(X_n,Y_n)$ be the observed data
and recall that the uni-modal regression finds the parameters using 
\begin{equation}
\begin{aligned}
\hat{\beta}_0, \hat{\beta}_1 
& = \underset{\beta_0,\beta_1}{{\sf argmax}}\,\, \frac{1}{nh}\sum_{i=1}^n K\left(\frac{Y_i- \beta_0-\beta_1X_i}{h}\right).
%& =  \underset{\beta_0,\beta_1}{{\sf argmax}}\,\, \frac{1}{n} \sum_{i=1}^n \Psi_i(\beta_0,\beta_1),
\end{aligned}
\label{eq::linear_modal2}
\end{equation}
%where $ \Psi_i(\beta_0,\beta_1) = K\left(\frac{\beta_0+\beta_1X_i-Y_i}{h}\right)$.
Note that when we take $K(x) = K_2(x) = \frac{1}{2}I(|x|\leq 1)$, we obtain equation \eqref{eq::linear_modal}.

%In general, there is no closed-form of $\hat{\beta}_0, \hat{\beta}_1 $. 

Given an initial choice of parameters $\beta_0^{(0)},\beta_1^{(0)}$,
the EM algorithm iterates the following two steps until convergence ($t=1,2,\cdots$):
\begin{itemize}
\item {\bf E-step.} Given $\beta_0^{(t-1)},\beta_1^{(t-1)}$, compute the weights
$$
%\pi(i|\beta_0^{(t-1)},\beta_1^{(t-1)}) = \frac{\Psi_i(\beta_0^{(t-1)},\beta_1^{(t-1)})}{\sum_{j=1}^n \Psi_j(\beta_0^{(t-1)},\beta_1^{(t-1)})}
\pi\left(i|\beta_0^{(t-1)},\beta_1^{(t-1)}\right) = \frac{K\left(\frac{Y_i- \beta_0^{(t-1)}-\beta_1^{(t-1)}X_i}{h}\right)}{\sum_{j=1}^n K\left(\frac{Y_j- \beta_0^{(t-1)}-\beta_1^{(t-1)}X_j}{h}\right)}
$$
for each $i=1,\cdots, n$.

\item {\bf M-step.} Given the weights, update the parameters by 
\begin{align*}
%\hat{\beta}^{(t)}_0, \hat{\beta}^{(t)}_1 =  \underset{\beta_0,\beta_1}{{\sf argmax}}\,\, \frac{1}{n} \sum_{i=1}^n \pi(i|\beta_0^{(t-1)},\beta_1^{(t-1)})\log\Psi_i(\beta_0,\beta_1)
\hat{\beta}^{(t)}_0, \hat{\beta}^{(t)}_1 &=  \underset{\beta_0,\beta_1}{{\sf argmax}}\,\, \frac{1}{n} \sum_{i=1}^n \pi(i|\beta_0^{(t-1)},\beta_1^{(t-1)})\log K\left(\frac{Y_i- \beta_0-\beta_1X_i}{h}\right).
%& = (\mathbb{X}^T \mathbb{W}_{(t)} \mathbb{X}^T)^{-1} \mathbb{X}^T \mathbb{W}_{(t)} \mathbb{Y},
\end{align*}
\end{itemize}
%Note that the maximizer in the M step has a matrix expression

When the kernel function $K$ is a Gaussian, the M-step has a closed-form expression:
$$
\hat{\beta}^{(t)}_0, \hat{\beta}^{(t)}_1 = (\mathbb{X}^T \mathbb{W}_{(t)} \mathbb{X}^T)^{-1} \mathbb{X}^T \mathbb{W}_{(t)} \mathbb{Y},
$$
where $\mathbb{X}^T = ((1, X_1)^T,(1,X_2)^T,\cdots, (1,X_n)^T)$ is the transpose of the covariate matrix in regression problem
and $\mathbb{W}_{(t)}$ is an $n\times n$ diagonal matrix with elements $$
\pi(1|\beta_0^{(t-1)},\beta_1^{(t-1)}),\cdots,\pi(n|\beta_0^{(t-1)},\beta_1^{(t-1)})
$$
and $\mathbb{Y} = (Y_1,\cdots,Y_n)^T$ is the response vector.
This is because the problem reduces to a weighted least square estimator in linear regression. 
Thus, the updates can be done very quickly.

Note that the EM algorithm may stuck at the local optima \citep{yao2014new} so the choice 
of initial parameters is very important. 
In practice, we would recommend to rerun the EM algorithm with many different initial parameters
to avoid the problem of falling in a local maximum. 

The EM algorithm can be extended to nonparametric uni-modal regression
as well. 
See \cite{yao2012local} for an example of applying the EM algorithm
to find the uni-modal regression
using a local polynomial estimator.

\subsection{Meanshift Algorithm}

To illustrate the principle of the meanshift algorithm \citep{comaniciu2002mean,fukunaga1975estimation,cheng1995mean}, 
we return to the 1D toy example.
Suppose that we observe IID random samples $Z_1,\cdots,Z_n\sim q$. 
Let $\hat{q}$ be a KDE with a Gaussian kernel $K_G(x) = \frac{1}{\sqrt{2\pi}}e^{-x^2/2}$.
A powerful feature of the Gaussian kernel is that its nicely behave derivative:
$$
K'_G(x) = -x \cdot \frac{1}{\sqrt{2\pi}}e^{-x^2/2} = -x\cdot K_G(x).
$$
The derivative of the KDE is then
\begin{align*}
\hat{q}'(z) &= \frac{d}{dz} \frac{1}{nh}\sum_{i=1}^n K_G\left(\frac{Z_i-z}{h}\right)\\
& = \frac{1}{nh^3 }\sum_{i=1}^n (Z_i-z) K_G\left(\frac{Z_i-z}{h}\right)\\
& =  \frac{1}{nh^3 }\sum_{i=1}^n Z_i  K_G\left(\frac{Z_i-z}{h}\right) - \frac{z}{nh^3}\sum_{i=1}^n K_G\left(\frac{Z_i-z}{h}\right).
\end{align*}
Multiplying both sides by $nh^3$ and dividing them by $\sum_{i=1}^n K_G\left(\frac{Z_i-z}{h}\right)$, the above equation becomes
$$
\frac{nh^3}{\sum_{i=1}^n K_G\left(\frac{Z_i-z}{h}\right)} \cdot \hat{q}'(z) = \frac{\sum_{i=1}^n Z_i  K_G\left(\frac{Z_i-z}{h}\right)}{\sum_{i=1}^n   K_G\left(\frac{Z_i-z}{h}\right)} - z. 
$$
Rearranging this expression, we obtain
$$
\underbrace{z}_\text{current location} + \underbrace{\frac{nh^3}{\sum_{i=1}^n K_G\left(\frac{Z_i-z}{h}\right)} \cdot \hat{q}'(z)}_\text{gradient aescent} = \underbrace{\frac{\sum_{i=1}^n Z_i  K_G\left(\frac{Z_i-z}{h}\right)}{\sum_{i=1}^n   K_G\left(\frac{Z_i-z}{h}\right)}}_\text{next location}
$$
Namely, given a point $z$, the value of $\frac{\sum_{i=1}^n Z_i  K_G\left(\frac{Z_i-z}{h}\right)}{\sum_{i=1}^n   K_G\left(\frac{Z_i-z}{h}\right)}$
is a shifted location by applying a gradient ascent with amount $\frac{nh^3}{\sum_{i=1}^n K_G\left(\frac{Z_i-z}{h}\right)} \cdot \hat{q}'(z)$.
Therefore, the meanshift algorithm updates an initial point $z^{(t)}$ as
$$
z^{(t+1)} = \frac{\sum_{i=1}^n Z_i  K_G\left(\frac{Z_i-z^{(t)}}{h}\right)}{\sum_{i=1}^n   K_G\left(\frac{Z_i-z^{(t)}}{h}\right)} 
$$
for $t=0,1,\cdots$. 
According to the above derivation, this update moves points by
a gradient ascent. 
Thus, the stationary point $z^{(\infty)}$ will one of the local modes of the KDE.
Note that although some initial points do not converge to a local modes,
these points forms a set with $0$ Lebesgue measure,
so can be ignored \citep{chen2017statistical}.

To generalize the meanshift algorithm to multi-modal regression, we fix the covariate
and shift only the response variable.
More specifically, given a pair of point $(x,y^{(0)})$, we
fix the covariate value $x$ and update the response variable as follows:
\begin{equation}
y^{(t+1)} = \frac{\sum_{i=1}^nY_i K_1\left(\frac{X_i-x}{h_1}\right)K_2\left(\frac{Y_i-y^{(t)}}{h_2}\right)}{\sum_{i=1}^n K_1\left(\frac{X_i-x}{h_1}\right)K_2\left(\frac{Y_i-y^{(t)}}{h_2}\right)},
\label{eq::PMC}
\end{equation}
for $t=0,1,\cdots$.
Here
$K_2=K_G$ is the Gaussian kernel although
the meanshift algorithm accomodates other kernel functions; see \cite{comaniciu2002mean} for a discussion.
The update in equation \eqref{eq::PMC} is called the conditional meanshift algorithm in \cite{Einbeck2006}
and the partial meanshift algorithm in \cite{chen2016nonparametric}. 
The conditional local modes include the
stationary points $y^{(\infty)}$.
To find all conditional local modes, we often start with multiple initial locations of the response variable
and apply equation \eqref{eq::PMC} to each of them. 

The kernel function for the covariate $K_1$ in equation \eqref{eq::PMC} 
is not limited to a Gaussian kernel.
$K_1$ can even be a deconvolution kernel 
in the presence of measurement errors \citep{zhou2016nonparametric}.

\subsection{Available Softwares}

There are many statistical packages in R for modal regression. 
On \texttt{CRAN}, there are two packages that contain functions for modal regression:
\begin{itemize}
\item \texttt{hdrcde}: the function \texttt{modalreg} computes a multi-modal regression using the method of \cite{Einbeck2006}. 
\item \texttt{lpme}: the function \texttt{modereg} performs a multi-modal regression
that can be used in situations
with or without measurement errors. This
package is based on the methods in \cite{zhou2016nonparametric}. 
Moreover, it also has two functions for bandwidth selection -- \texttt{moderegbw} and \texttt{moderegbwSIMEX} --
that apply the bandwidth selectors described in \cite{zhou2017bandwidth} and \cite{zhou2016nonparametric}. 
\end{itemize}
Note that there is also an R package on github for modal regression: \url{https://github.com/yenchic/ModalRegression}.
This package is based on the method of \cite{chen2016nonparametric}. 

\section{Similar Approaches to Modal Regression}	\label{sec::similar}

Multi-modal regression is a powerful tool for detecting multiple
components of the conditional density. 
%In particular, when the response variable and the covariate are associated
%in a complicated ways such as having multiple components, we may observe multiple components in the scatter plot. 
The right panel of Figure~\ref{fig::ex02} demonstrates the the power of a compact prediction set
obtained by multi-modal regression.
%Moreover, the multi-modal regression allows us to detect the hidden structures inside the data
%without specifying any particular form of the relationship between $X$ and $Y$. 
Multiple components of the conditional density can also be obtained by
other approaches such as the mixture of regression
and regression quantization method.
These approaches are briefly described below.

\subsection{Mixture of Regression}
When multiple components reside in the conditional density, 
traditional regression analysis applies a mixture of regression model \citep{quandt1972new,quandt1978estimating,lindsay1995mixture}.
A general form of a mixture of regression model \citep{Huang2012jasa,Huang2013jasa} is
$$
Y|X=x \sim \sum_{\ell=1}^L \pi_\ell(x) N(m_\ell(x), \sigma_\ell^2(x)),
$$
where $\pi_\ell(x)\geq0$ is the proportion of the $\ell$-th component (note that  $\sum_{\ell=1}^L \pi_\ell(x) = 1$)
and $m_\ell(x),\sigma^2_\ell(x)$ denote the mean and variance.
Here we assume that the data comprises $L$ mixtures of Gaussian components
(note that this assumption can be relaxed as well). 
In this case, the parameter functions $\pi_\ell(x)$, $m_\ell(x)$, and $\sigma_\ell^2(x)$
are parameters of interest
and must be estimated from the data.
The parameter functions can be estimated using smoothing techniques and maximum likelihood estimation \cite{Huang2013jasa}.

Although the mixture of regression approach is flexible, it has several limitations.
First is the identifiability problem; different combinations of the parameter functions may lead to the same or similar
conditional density, which destabilizes the estimator. 
Second, the number of components $L$, must be known a priori. 
If we assume a parametric model for the parameter functions,
$L$ can be chosen by a model selection criterion 
such as the Akaike or Bayesian information criterion \citep{Huang2013jasa}. 
However, assuming a parametric form decreases the flexibility of the model.
Moreover, computing the estimators of parameter functions
often requires an EM-algorithm, which may need several re-initializations of the initial condition 
to get a desired estimate.

\subsection{Regression Quantization}
Alternatively,
the authors of \cite{loubes2017prediction} detected multiple components in a conditional density function
by combining $k$-means (vector quantization; \citealt{graf2007foundations,gersho2012vector}) algorithm
and $k$-nearest neighbor ($k$NN) approach.
Their method is called regression quantization. 
To illustrate the idea, we consider a $1D$ Gaussian mixture model with $L$ distinct components. 
If the components are well-separated and their proportions are similar, the a $k$-means algorithm with $k=L$ will return
$k$ points (called centers in the $k$-means literature) that approximate the centers of Gaussians. 
Thus, the centers of $k$-means correspond to the centers of components in our data.
%In more detail, let $Z_1,\cdots,Z_n$ be 1D random variables. 
%The $k$-means algorithm finds $k$ points $c_1,\cdots,c_k$ such that
%$$
%\Phi_n(c_1,\cdots,c_k) = \sum_{i=1}^n\min_{j=1,\cdots,k}\|Z_i-c_j\|^2
%$$
%is minimized. 

In a regression setting,
the $k$-means algorithm is combined with $k$NN.
To avoid conflict in the notations, we denote the number of centers in the $k$-means by $L$,
although the algorithm itself is called $k$-means. 
For a given point $x$, we find those $X_i$'s within the $k$-nearest neighborhood of $x$,
process their corresponding responses by the $k$-means algorithm.
Let 
$$
W_{n,i}(x) = \begin{cases}
\frac{1}{k} \quad &\mbox{if $X_i$ is among the $k$ nearest neighbor of $x$} \\
0 \quad &\mbox{otherwise}
\end{cases}
$$
be the weight of each observation. 
Given a point $x$, the estimator in \cite{loubes2017prediction} was defined as
$$
\hat{c}_1(x),\cdots,\hat{c}_L(x) = \underset{c_1,\cdots,c_L}{\sf argmin} \sum_{i=1}^n\min_{j=1,\cdots,L}W_{n,i}(x)\|Y_i-c_j\|^2.
$$
Namely, we apply the $k$-means algorithm to the response variable of
the $k$-NN observations. 
For correct choices of $k$ and $L$, the resulting estimators
properly summarize the data.
Because $k$ behaves like the smoothing parameter in the KDE, 
the choice of $k=k_n$ has been theoretically analyzed \citep{loubes2017prediction}. 
However, the choice of $L$ often relies on
prior knowledge about the data (number of components), although $L$ can be chosen by
a gap heuristic approach \citep{tibshirani2001estimating}.

%In the case like Figure xxx, 
%there are some other approaches that have been proposed
%to analyze data. 
%The first approach is a mixture of regression approach and the other is a quantization approach.
%
%
%quantization method \cite{loubes2017prediction}
%
%mixture model \cite{lindsay1995mixture}
%\cite{Huang2012jasa,Huang2013jasa}
%

\section{Discussion}	\label{sec::future}

%Compared to the usual regression model (conditional expectation) and
%the quantile regression, the study in modal regression
%has attracted less attention.

This paper reviewed common methods for fitting modal regressions. 
We discussed both uni-modal and multi-modal approaches, along with
relevant topics such as large sample theories, 
bandwidth selectors, and computational recipes. 
Here we outline some possible future directions
of modal regression.
\begin{itemize}

\item {\bf Multi-modal regression in complex processes.}
Although the behavior of uni-modal regression has been analyzed 
in dependent and censored scenarios
\citep{collomb1986note,ould2005strong,khardani2010some,khardani2011uniform}, 
the behavior of multi-modal regression is still unclear.
%So an open question is to investigate
The behavior of multi-modal regression in censoring cases remains an open question.
Moreover,
by
comparing the estimator in censored response variable cases and
measurement error cases (equation \eqref{eq::censor} and \eqref{eq::de_KDE}, respectively), 
we find that to account for censoring,
we must change the kernel function of the response variable,
whereas to adjust the
measurement errors ,
we need to modify the kernel function of the covariate. 
Thus, we believe that the KDE
can be modified 
to solve
the censoring and measurement error problems
at the same time.

\item {\bf Valid confidence band.}
In \cite{chen2016nonparametric}, the
confidence band of the multi-modal regression was constructed by a bootstrap approach.
However, because this confidence band does not correct bias in KDE,
it requires an
undersmoothing assumption. 
Recently, \cite{calonico2017effect} proposed a debiased approach that
constructs a bootstrap nonparametric confidence set without undersmoothing. %for density estimation and usual regression analysis.
The application of this approach to modal regression is another possible future direction.

\item {\bf Conditional bump hunting.}
A classical problem in nonparametric statistics is bump hunting \citep{good1980density, hall2004bump,burman2009multivariate},
which detects the number of significant local modes. 
In modal regression analysis, 
the bump hunting problem may be studied in a regression setting.
%we may study the bump hunting problem in a regression setting.
%bump hunting can be applied in a regression setting. 
More specifically, we want to detect the number of significant local modes
of the conditional density function. 
The output will be an integer function of the covariate 
that informs how the number of significant local modes changes
over different covariate values.

%\item {\bf High dimensional modal regression.}
%Studies in modal regression have been focusing on a low dimensional situation.
%Namely, the dimension of covariate is fixed and is less than the sample size $n$. 
%When the dimension is large, or even in a high dimensional regime where $p>> n$, 
%how to perform a modal regression is still unclear. 
%A possible approach is to propose a linear model and use equation \eqref{eq::argmin}
%with a penalty to frame the problem as a minimization problem such as using a LASSO estimator:
%$$
%\hat{\mathbb{\beta}} = \underset{\beta}{\sf argmin} \,\, \frac{1}{2nh}\sum_{i=1}^n I\left(|Y_i-{\beta}^T X_i|> h\right)
%+ \lambda|\hat{\mathbb{\beta}}|_1,
%$$
%where ${\mathbb{\beta}} = ({\beta}_0,{\beta}_1,\cdots,{\beta}_p)$
%and $|\cdot|_1$ is the $L_1$ norm of a vector and $\lambda$ is a tuning parameter.
%This provides not only an estimate of parameters 
%but also performs a variable selection. 
%The choice of $h$ and $\lambda$ can then be done by cross-validation. 

%\item {\bf Bandwidth selection. }

\end{itemize}

\section*{Acknowledgement}
We thank two referees and the editor for their very helpful comments. 
Yen-Chi Chen is supported by NIH grant number U01 AG016976. 

\pagebreak

\bibliographystyle{apa}
\bibliography{modal_review}

%\subsection*{\sffamily \Large FURTHER READING}
%%Please insert any further reading/resources here.
%For readers who may want more information on concepts in your article, provide full references and/or links to additional recommended resources (books, articles, websites, videos, datasets, etc.) that are not included in the reference section. Please do not include links to non-academic sites, such as Wikipedia, or to impermanent websites.

\end{document}